\documentclass[conference]{IEEEtran}
\usepackage{cite}
\usepackage{amsmath,amssymb,amsfonts}
\usepackage{algorithmic}
\usepackage{graphicx}
\usepackage{textcomp}
\usepackage{booktabs}
\usepackage{url}
\usepackage{xcolor}
\usepackage{array}
\usepackage{multirow}
\usepackage{makecell}

\def\BibTeX{{\rm B\kern-.05em{\sc i\kern-.025em b}\kern-.08em
    T\kern-.1667em\lower.7ex\hbox{E}\kern-.125emX}}

\begin{document}

\title{Comparative Analysis of Direct-to-Cell (D2C) and 3GPP Non-Terrestrial Networks (NTN) for Global Connectivity}

\author{
    \IEEEauthorblockN{Donglin Wang\IEEEauthorrefmark{2},   Anjie Qiu\IEEEauthorrefmark{2}, Qiuheng Zhou\IEEEauthorrefmark{1}, and Hans D. Schotten\IEEEauthorrefmark{2}\IEEEauthorrefmark{1}}
    \IEEEauthorblockA{\textit{\IEEEauthorrefmark{2}Rhineland-Palatinate Technical University of Kaiserslautern-Landau, Germany} \\
        \{dwang, qiu, schotten\}@eit.uni-kl.de}
    \IEEEauthorblockA{\textit{\IEEEauthorrefmark{1}German Research Center for Artificial Intelligence (DFKI GmbH), Kaiserslautern, Germany} \\
        \{qiuheng.zhou, schotten\}@dfki.de}
}

\maketitle

\begin{abstract}
The quest for ubiquitous mobile coverage has catalyzed two fundamentally distinct architectural paradigms: Direct-to-Cell (D2C) and standardized 3GPP Non-Terrestrial Networks (NTN). D2C, pioneered by SpaceX Starlink and AST SpaceMobile, leverages existing terrestrial spectrum and unmodified consumer handsets to provide emergency connectivity as a market-driven overlay. In contrast, 3GPP NTN, standardized across Releases 17–19, offers a systematic, satellite-native framework designed for long-term scalability, high-throughput broadband, and deep integration with terrestrial 5G/6G networks. This paper provides a comprehensive technical comparison of these approaches, analyzing their standardization trajectories, network architectures, physical layer innovations, security postures, and operational trade-offs. We further examine their implications for emerging 6G use cases, particularly autonomous driving, where safety-critical redundancy demands a hybrid tri-link architecture combining terrestrial 5G, NTN broadband, and D2C emergency fallback. Our analysis reveals that while D2C enables rapid market entry with legacy device compatibility, NTN delivers superior performance, security, and scalability, positioning it as the foundational framework for 6G satellite-terrestrial convergence. A hybrid model leveraging the strengths of both paradigms is identified as the most viable path toward true global connectivity.
\end{abstract}

\begin{IEEEkeywords}
5G, 6G, Terrestrial Networks (TN), Non-Terrestrial Networks (NTN), Direct-to-Cell (D2C), Satellite Communications, 3GPP Release 17/18/19, Autonomous Driving, Vehicle-to-Everything (V2X) Communications
\end{IEEEkeywords}

\section{Introduction}

\subsection{Motivation and Background}

Global cellular coverage remains fragmented: terrestrial networks (TNs) provide coverage to approximately 80\% of the world's population, yet over 70\% of Earth's landmass remains beyond the reach of conventional mobile infrastructure \cite{GSMA_2024, WorldBank_2024}. This stark connectivity gap creates critical challenges across emergency services, maritime operations, remote industries (agriculture, forestry, mining), and next-generation autonomous mobility systems \cite{graydon2020connecting, sun2025investigation}. Achieving ubiquitous global connectivity is therefore a fundamental imperative for 5G/6G evolution and a key component of sustainable development objectives \cite{chowdhury20206g, Giordani_2020}.

\begin{figure}[htp]
\centering
\includegraphics[width=\columnwidth]{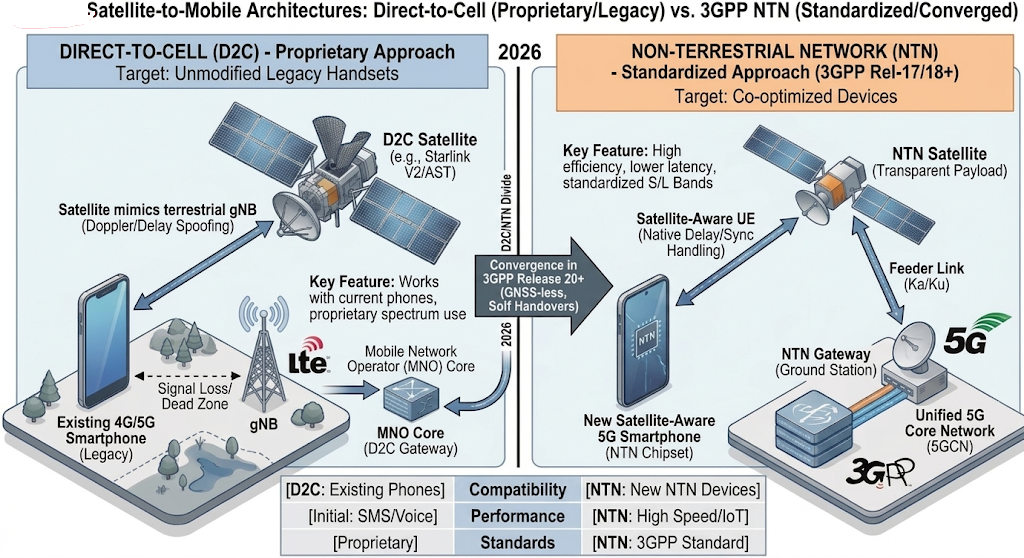}
\caption{Comparative Analysis of D2C and Standardized 3GPP NTN Architectures.}
\label{fig:dtc_vs_ntn}
\end{figure}

Two fundamentally distinct technological paradigms have emerged to address this gap over the period 2022--2026, as shown in Fig. \ref{fig:dtc_vs_ntn}: \textbf{Direct-to-Cell (D2C)}, a market-driven, spectrum-opportunistic approach utilizing terrestrial bands to provide immediate connectivity to unmodified 4G/5G handsets, prioritizing rapid time-to-market and universal device compatibility \cite{spacex_tmobile_2024, AST_2024}; and \textbf{3GPP Non-Terrestrial Networks (NTN)}, a standards-based framework evolved through Releases 17--19 that uses dedicated satellite spectrum, supports transparent and regenerative architectures, and targets long-term scalability and deep integration with terrestrial 5G New Radio (NR) and Internet of Things (IoT) services \cite{3gpp_ts_38_300_2026, heo2023mimo}.


Both paradigms represent legitimate but fundamentally different engineering philosophies: D2C emphasizes ecosystem compatibility and time-to-market; NTN emphasizes performance, standardization, and long-term convergence. Understanding their respective trade-offs is critical for standardization bodies, regulators, operators, and automotive original equipment manufacturers (OEMs) making technology investment decisions through 2030.

\subsection{State-of-the-Art and Related Work}

Satellite-terrestrial integration research spans multiple domains. Heo et al. \cite{heo2023mimo} and Zong et al. \cite{chowdhury20206g} provide comprehensive surveys of physical layer technologies; Samsung \cite{Samsung_2024} and Giordani et al. \cite{Giordani_2020} position satellite integration as foundational to 6G vision. The 3GPP standardization trajectory progressed through Release 17 feasibility studies (2022), Release 18's transparent payload specification with Doppler pre-compensation (2024) \cite{3gpp_ts_38_104_2024}, and Release 19's regenerative payload support with inter-satellite links (ISLs) and AI-optimized beam scheduling (2026) \cite{3gpp_ts_38_300_2026}. These works form the context for NTN technical foundations documented in this paper.

Parallel to standards-based NTN, SpaceX Starlink deployed operational D2C emergency messaging (T-Mobile partnership, 2024), followed by Globalstar and Iridium implementations. The Federal Communications Commission (FCC)'s supplemental coverage from space (SCS) regulatory framework \cite{FCC_2024} permits secondary satellite operation in terrestrial personal communications service (PCS), advanced wireless services (AWS) bands, enabling rapid deployment without 3GPP standardization cycles. However, this market-driven approach introduces unresolved interference coordination challenges and security vulnerabilities. Recent analysis—Liu et al. \cite{leod2c2025} identifying location-leakage attacks via timing advance side-channels—highlights D2C security gaps absent in standards-based NTN. Additionally, D2C's nascent state (SMS-only in 2024) contrasts sharply with NTN's multi-release evolutionary path, creating distinct technical trade-offs examined herein.



Vehicle-to-everything (V2X) communication over satellite backhaul represents an emerging research frontier. SAE J3016 standardization \cite{SAE_J3016_2021} defines five levels of autonomous driving automation, with Level 4 and 5 implementations (Waymo, Tesla, Mercedes, BMW) requiring safety-critical redundancy across multiple connectivity technologies. Recent work by Scholz et al. \cite{wang2025survey} analyzes V2X standardization challenges for autonomous vehicles, emphasizing the role of hybrid satellite-terrestrial architectures. Automotive OEM roadmaps (Tesla Cybertruck, BMW iX M60, Waymo Robotaxi, Mercedes Drive Pilot) increasingly incorporate satellite fallback mechanisms, validating the commercial relevance of satellite-based vehicle connectivity.

While existing NTN surveys (\cite{heo2023mimo, chowdhury20206g}) focus on satellite-terrestrial convergence and \cite{Giordani_2020} on 6G architecture, this paper uniquely provides a direct D2C vs. NTN technical comparison across standardization, structure, and application—a gap in published literature as of 2026.

\subsection{Paper Organization}

The paper is organized as follows: Section II examines standardization and regulatory frameworks. Section III analyzes the D2C network structure, the NTN antenna payload architectures, and the Tri-link architecture design. Section IV presents technical comparisons, including Doppler management and performance metrics. Section V explores the applications and use cases of the paradigms. Section VI discusses the D2C security implications. Section VII discusses 6G evolution pathways. Section VIII concludes with a convergence outlook and future directions.

\section{Standards and Regulatory Framework}
\subsection{Direct-to-Cell (D2C)}

D2C represents a departure from traditional telecommunications standardization, instead pursuing a rapid, market-driven approach. D2C operators (SpaceX Starlink, AST BlueWalker 3, Iridium, Globalstar) lease terrestrial mobile spectrum under the FCC's SCS framework (2024), where mobile network operators (MNOs) retain primary access and satellites act as secondary service providers; SpaceX, for example, uses T-Mobile's PCS (1.9 GHz) and AWS (1.7/2.1 GHz) bands \cite{spacex_tmobile_2024}. This model is further constrained by geographic exclusion zones, stringent power flux density (PFD) limits, and International Telecommunication Union (ITU) Radio Regulations Article 22, which creates substantial coordination barriers for international deployment \cite{ITU_RR_2024}.
\subsection{3GPP Non-Terrestrial Networks (NTN)}

In contrast, NTN follows a systematic, multi-release standardization trajectory:

\subsubsection{Release Timeline and Milestones  as shown in Table \ref{tab:ntn_standardization}}

\begin{table*}[htbp]
\caption{3GPP NTN Standardization Timeline}
\label{tab:ntn_standardization}
\centering
\begin{tabular}{|l|l|p{8cm}|}
\hline
\textbf{Release} & \textbf{Date} & \textbf{Key Features} \\
\hline
Release 17 & Dec 2022 & Initial study item; feasibility assessment; channel models; architecture options; identification of necessary enhancements \cite{3gpp38821} \\
\hline
Release 18 & Mar 2024 & First complete specification; transparent payload support; NB-IoT/enhanced Machine-Type Communication (eMTC) over NTN; frequency bands $n255/n256$ (S-band); Doppler pre-compensation \cite{3gpp_ts_38_104_2024} \\
\hline
Release 19 & Mar 2026 & Regenerative payload support; ISLs; enhanced mobility; AI-driven beam scheduling; multi-connectivity \cite{3gpp_ts_38_300_2026} \\
\hline
Release 20 & Planned 2028 & Full 6G NTN integration; native AI/ML interfaces; terahertz band support; ubiquitous positioning; D2C-to-NTN interoperability \\
\hline
\end{tabular}
\end{table*}

\subsubsection{Spectrum Allocation}

NTN operates in dedicated mobile satellite service (MSS) bands, ensuring separation from TNs :

\begin{itemize}
    \item S-band (n255/n256): 1980-2010 MHz (uplink), 2170-2200 MHz (downlink) are primary bands for Release 18 NTN
    \item L-band (n256): 1525-1559 MHz (downlink), 1626.5-1660.5 MHz (uplink) are complementary bands for IoT services
    \item Ka-band (n510/n511): 27.5-30.0 GHz (uplink), 17.7-20.2 GHz (downlink) are planned for Release 19/20 high-throughput services
\end{itemize}

This dedicated spectrum allocation eliminates coexistence challenges with TNs and provides predictable interference environments.

\subsubsection{Regulatory Harmonization}

NTN benefits from a more harmonized regulatory environment through ITU-recognized MSS allocations, 3GPP-defined interoperable frequency bands, established coordination mechanisms at the national level, and a clear roadmap spanning  World Radiocommunication Conference (WRC)-23 and WRC-27 agenda items for NTN-specific spectrum needs \cite{ITU_WRC23_FinalActs,ITU_AI1_11_WRC27}.

\section{Network Structures and Payload Architectures}

\subsection{D2C Architecture: The "Space Tower" Model}
The D2C structure, often called the "Space Tower" model, operates as a transparent extension of TNs. Each satellite acts as a high-altitude evolved Node B (eNB) or gNode B (gNB), relying on massive phased-array antennas (30--64 m$^2$) and onboard Doppler pre-compensation, timing advance calculation, and power-control signaling to support unmodified 4G/5G handsets over $\sim$180 dB L-band path loss. Backhaul via X/Ka-band feeder links connects to terrestrial gateways, and Starlink's ISLs can reduce latency to 50--100 ms; however, D2C still faces lower link margins (7.2 dB vs. 10.2 dB for NTN), 150--250 ms gateway-routing latency, uplink interference from millions of legacy devices, and inter-satellite footprint coverage gaps.

\subsection{NTN Architecture}

NTN is an end-to-end (E2E) satellite-terrestrial access architecture. It comprises NTN-capable user equipment (UE), a service link between the UE and satellite, a feeder link between the satellite and gateway, and integration with terrestrial gNB and 5G core functions. In LEO operation, the satellite forms a moving cell, so beam steering, timing advance, Doppler assistance, and ephemeris-aware synchronization are architectural requirements rather than optional optimizations.

Within this system, 3GPP standardizes two complementary payload models. \textbf{Transparent payloads} implement radio frequency-in/radio frequency-out (RF-in/RF-out) relay with most base-station processing anchored on the ground, which simplifies the spacecraft and enables fast software updates but increases gateway dependence and transport latency. \textbf{Regenerative payloads} move selected PHY and MAC functions onboard so the satellite behaves more like a distributed gNB, supporting local scheduling, ISLs, mesh routing, and multi-connectivity with latency as low as 25 ms. The trade-off is higher satellite mass, power consumption, and onboard complexity in exchange for lower latency, reduced gateway reliance, and stronger 6G scalability. Table \ref{tab:ntn_payload_comparison} summarizes the main differences across Releases 17--19.

\begin{table*}[htbp]
\caption{Comparison of NTN Payload Architectures}
\label{tab:ntn_payload_comparison}
\centering
\begin{tabular}{|l|p{4cm}|p{4cm}|}
\hline
\textbf{Characteristic} & \textbf{Transparent Payload} & \textbf{Regenerative Payload} \\
\hline
Processing Location & Ground-based gNB & Onboard satellite \\
\hline
Latency (LEO, one-way) & $\sim 60$ ms & $\sim 25$ ms (with ISLs) \\
\hline
Satellite Complexity & Low & High (2–3$\times$) \\
\hline
Inter-Satellite Links & Not practical & Supported \\
\hline
Gateway Dependence & High (every satellite pass) & Reduced (ISL routing) \\
\hline
Protocol Agility & High (ground software updates) & Moderate (onboard software) \\
\hline
Cost per Satellite & Lower & Higher \\
\hline
6G Scalability & Limited & High \\
\hline
\end{tabular}
\end{table*}

\subsection{Tri-Link Architecture Design}

\subsubsection{Motivation and Three-Tier Decomposition}

\begin{figure}[htp]
\centering
\includegraphics[width=\columnwidth]{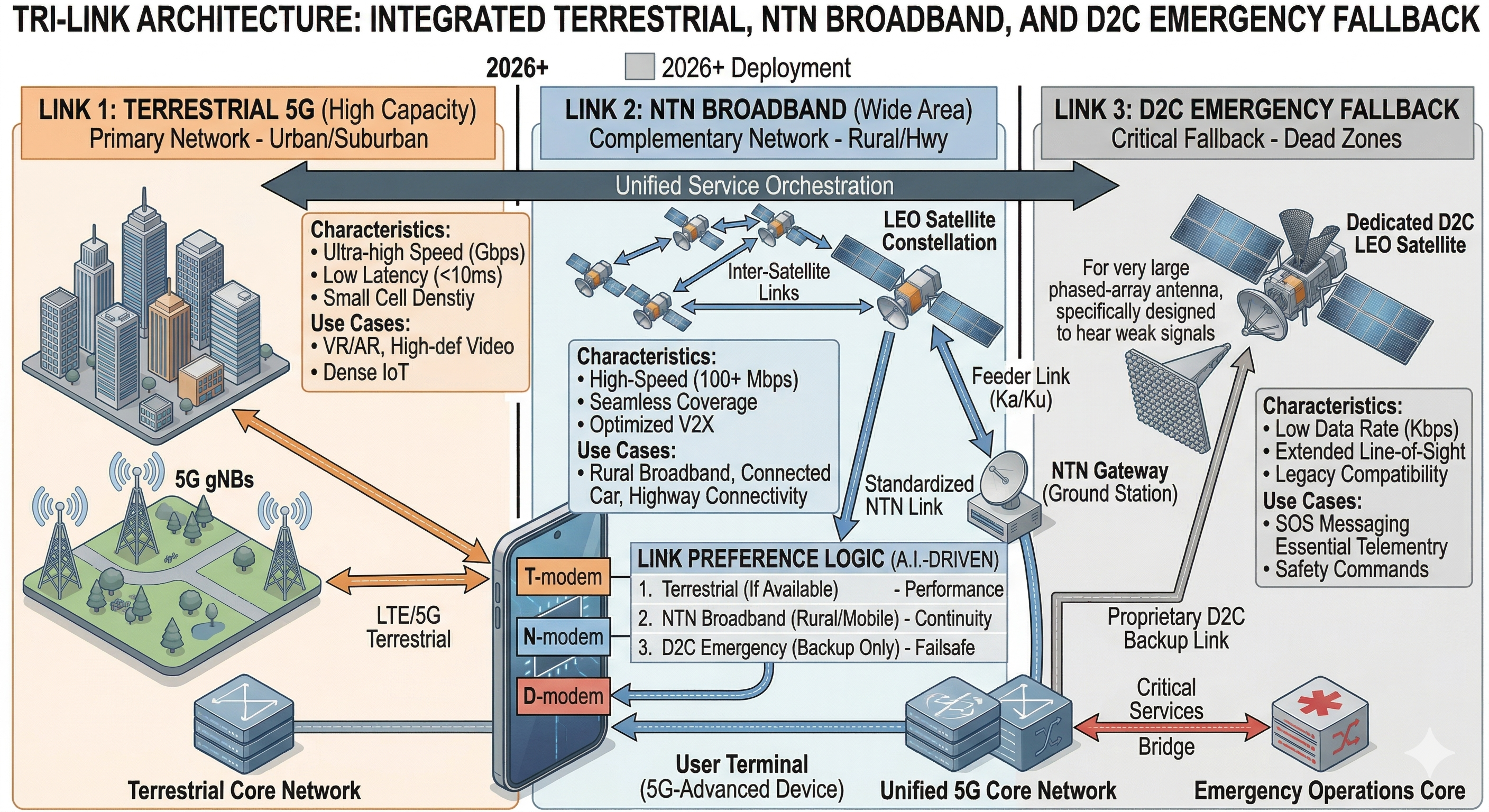}
\caption{Tri-Link Architecture for Ubiquitous and Resilient Connectivity (2026+).}
\label{fig:trilink}
\end{figure}

No single wireless paradigm achieves simultaneously (i) low latency, (ii) high bandwidth, (iii) ubiquitous coverage, and (iv) deterministic availability. The tri-link architecture assigns these conflicting requirements across three tiers as illustrated in Fig. \ref{fig:trilink}:

\begin{itemize}
    \item \textbf{Tier 1 — Terrestrial 5G}: \emph{latency} $< 20$ ms, \emph{throughput} $> 100$ Mbps, \emph{availability} 99.5\% (urban/suburban, ~80\% coverage). Primary role: real-time high-capacity services.
    
    \item \textbf{Tier 2 — NTN Satellite}: \emph{latency} 30--60 ms, \emph{throughput} 20--50 Mbps, \emph{availability} 95\% globally (~15\% coverage in rural/maritime). Primary role: bridging terrestrial coverage gaps.
    
    \item \textbf{Tier 3 — D2C Fallback}: \emph{latency} 150--250 ms, \emph{throughput} 1--10 kbps, \emph{availability} 85\% (~5\% edge cases). Primary role: emergency messaging when Tiers 1--2 fail.
\end{itemize}

\subsubsection{Link Selection Logic and Handover Semantics}

Deterministic priority-based link selection routes traffic via: (1) \textbf{Tier 1 Available}: all traffic routed via 5G with continuous RSSI monitoring; (2) \textbf{Tier 1 Degraded} (RSSI $< -110$ dBm): latency-critical functions maintain Tier 1; bandwidth services transition to Tier 2; (3) \textbf{Tier 1 Lost}: failover to Tier 2 (NTN, 30--60 ms latency); (4) \textbf{Tiers 1 \& 2 Lost}: move to Tier 3 emergency channel (1--10 kbps).

Release 19 standardizes handover properties: latency visibility (transitions proceed only if next-tier latency meets requirements), deterministic state transitions (time-critical operations frozen during handover: $< 150$ ms within-tier, $< 500$ ms between-tier), explicit latency signaling (pre-computed fallback trajectories), and cryptographic tokens preventing unauthorized switching.

\subsubsection{Multi-Connectivity and Global Resilience}

Triple-tier redundancy enables service continuity during single-tier failure. Release 18/19 multi-connectivity to 2--3 concurrent NTN satellites aggregates 50--100 Mbps downlink, supporting bandwidth-intensive services. Message routing is tier-agnostic: endpoints transmit via Tier 1 while receiving relay traffic via Tier 2 satellite backhaul. Release 20 standardized interoperability enables seamless operator transition (OneWeb, Kuiper, Starshield) without service interruption, preventing single-operator monopoly failure.

\section{Technical Comparison and Trade-offs}


\subsection{Doppler Shift Management}

Both D2C and 3GPP NTN frameworks must address severe Doppler shifts caused by LEO satellite motion, distinguishing between instantaneous frequency shifts and frequency drift rates, both of which challenge frequency synchronization in receiver chains. Table \ref{tab:doppler_params} summarizes Doppler-related parameters for LEO satellite operations. The instantaneous Doppler frequency shift depends on the satellite's geometric position relative to the UE:

\begin{equation}
f_d(t) = \frac{f_c \cdot v_{sat} \cdot \cos(\theta(t))}{c}
\end{equation}

where $f_c$ is the carrier frequency (L-band: $\sim 1.6$ GHz for D2C/NTN), $v_{sat}$ is the orbital tangential velocity ($\approx 7.5$ km/s for LEO), $\theta(t)$ is the instantaneous angle between satellite velocity vector and line-of-sight (LOS), and $c$ is the speed of light.










\subsection{D2C vs. NTN Doppler Shift Compensation Strategies}

\begin{itemize}
    
    \item \textbf{D2C Satellite-Side Pre-Compensation Approach}: The satellite pre-calculates expected Doppler for each user based on ephemeris data and pre-distorts all downlink transmissions. Unmodified handsets receive already-compensated signals, imposing no firmware changes. However, pre-compensation introduces $(1\text{--}3)$ dB link margin penalty due to: (i) ephemeris prediction errors ($\sim \pm 2$ km), (ii) user motion perpendicular to LOS, (iii) atmospheric refraction. This approach sacrifices performance for universal compatibility.
    
    \item \textbf{NTN Joint Satellite-UE Compensation Approach}: NTN-capable UEs perform onboard Doppler estimation and compensation. Release 18 standardizes satellite-provided assistance data (ephemeris + satellite velocity vector), enabling UE-side frequency tracking loops with faster convergence ($\sim 100$ ms) and robust tracking of Doppler rates. This dual strategy achieves $(2\text{--}3)$ dB better link margin than D2C, critical for power-limited handsets \cite{3gpp_ts_38_104_2024}.
\end{itemize}

\begin{table}[htbp]
\caption{Doppler Shift and Rate Parameters: LEO Satellite Communication.}
\label{tab:doppler_params}
\centering
\begin{tabular}{|l|c|c|}
\hline
\textbf{Parameter} & \textbf{Symbol} & \textbf{Typical LEO Value} \\
\hline
Carrier Frequency (L-band) & $f_c$ & $1.5\text{--}2.5$ GHz \\
\hline
Orbital Velocity & $v_{sat}$ & $\approx 7.5$ km/s \\
\hline
Maximum Doppler Shift & $f_{d,max}$ & $\pm 26\text{--}40$ kHz \\
\hline
Doppler Rate (at Zenith) & $\dot{f}_d$ & $\approx 0.5\text{--}1$ kHz/s \\
\hline
Slant Range & $d$ & $400\text{--}2000$ km \\
\hline
Handover Re-sync Latency & $\tau$ & \makecell[c]{D2C: 150+ ms;\\ NTN: 50–100 ms}   \\
\hline
\end{tabular}
\end{table}
\subsection{Impact on Synchronization Design}
Unlike TNs where Doppler shifts are negligible ($< 100$ Hz at vehicular speeds), LEO satellite systems must maintain frequency tracking loops capable of (i) acquiring frequency offsets up to $\pm 40$ kHz, (ii) tracking rates of $\pm 1$ kHz/s, and (iii) maintaining coherence across $50$--$150$ ms handover transitions. 

The choice between D2C and NTN compensation philosophies directly determines receiver synchronization complexity and performance margins. D2C's satellite-side pre-compensation simplifies handset firmware (legacy devices need no modifications) but leaves residual frequency errors ($\pm 0.5$--$1$ kHz) that degrade coherent demodulation, translating to the documented $1$--$3$ dB link margin penalty. In contrast, NTN's joint compensation architecture distributes tracking burdens between satellite (coarse assistance) and UE (fine estimation), achieving tighter residual error budgets ($\pm 0.2$ kHz) critical for power-limited IoT and safety-critical automotive applications. NTN's superior Doppler handling, combined with satellite-provided synchronization assistance and Release 18's standardized tracking algorithms, enables safety-critical autonomous vehicle V2X operations requiring phase coherence ($< 0.1$ rad/symbol) across $50$--$150$ ms handover transitions without state loss.



    

\subsection{Timing Advance Management}

LEO propagation delays vary rapidly as satellites cross the sky, requiring continuous adjustment. Synchronization latency refers to the time required for initial acquisition or re-synchronization after handover transitions, typically 50–150 ms in LEO systems.

D2C calculates timing advance onboard and broadcasts it through standard LTE/5G procedures, but the rapid variation ($\sim 20$ $\mu$s/s at L-band) challenges legacy tracking loops. NTN Release 18 introduces higher update rates and predictive timing advance based on satellite ephemeris, reducing re-synchronization time to roughly 100 ms versus 150+ ms for D2C while preserving coherence at low elevations \cite{3gpp_ts_38_104_2024}.

These timing and waveform differences contribute directly to uplink performance. NTN's higher uplink margin stems from UE antenna gain ($+2$ dBi vs. $-2$ dBi for D2C) and optimized waveforms.

\subsection{Performance Characteristics}
Performance can be compared across data rate, latency, and power consumption. D2C currently supports SMS (150 bps, 2024), voice (10--50 kbps, 2026), and limited data expansion (100--500 kbps, 2028) \cite{d2cstarlink2025}; NTN Release 18 supports NB-IoT, LTE-M, and handheld 5G-NR, while Release 19/20 targets 20--100 Mbps. D2C gateway-routed latency is typically 150--250 ms round-trip time (RTT), reduced to 50--100 ms with ISL routing. In contrast, NTN reaches about 60 ms with transparent payloads and 25 ms with regenerative payloads and ISLs. On the power side, D2C handsets consume 5--6$\times$ terrestrial cellular power, while NTN-optimized terminals typically require 2--3$\times$ thanks to better link adaptation and ephemeris assistance. Table \ref{tab:comprehensive} summarizes these dimensions.

\begin{table*}[htbp]
\caption{Comprehensive Technical Comparison: D2C vs. 3GPP NTN}
\label{tab:comprehensive}
\centering
\begin{tabular}{|p{3cm}|p{5.5cm}|p{5.5cm}|}
\hline
\textbf{Feature} & \textbf{D2C} & \textbf{3GPP NTN} \\
\hline
Handset Compatibility & Unmodified 4G/5G legacy devices; no hardware changes required & NTN-aware firmware/chipset required (Release 17+); premium $\sim 10$–$15\%$ \\
\hline
Spectrum & Terrestrial bands (L-band PCS/AWS); shared with MNOs; interference risk & Dedicated MSS bands (S-band $n255/n256$); clean spectrum; regulatory separation \\
\hline
 Link Margin (Uplink) & $\sim 7$–$9$ dB (constrained by handset power) & $\sim 10$–$12$ dB (NTN-optimized receivers) \\
\hline
Data Rate & SMS (2024), voice (2026), limited data (2028) & IoT ($\sim 100$ kbps), voice ($\sim 50$ kbps), broadband ($\sim 20$–$100$ Mbps) \\
\hline
Latency & Gateway-routed: $\sim 150$–$250$ ms; ISL-routed: $\sim 50$–$100$ ms & Transparent: $\sim 60$ ms; Regenerative+ISL: $\sim 25$ ms \\
\hline
Battery Consumption & $5$–$6\times$ terrestrial & $2$–$3\times$ terrestrial \\
\hline
Security & Vulnerable: LTE evolved packet system (EPS) mobility management (EMM) spoofing; international mobile subscriber identity (IMSI) catcher; location leakage via TA side-channel & Enhanced: NTN-specific authentication; Doppler-robust temporary mobile subscriber identity (TMSI); ISL encryption \\
\hline
Coverage Pattern & Footprint overlaps; dense constellations minimize gaps & Planned geometry with deterministic coverage \\
\hline
Scalability to 6G & Limited (protocol stack frozen) & Extensible: AI-driven optimization; new waveforms OTFS, AFDM, OCDM \\
\hline
Deployment Status & Commercial SMS (2024); expanding & Release 18 (2024); Release 19 (2026) \\
\hline
\end{tabular}

\end{table*}

Table \ref{tab:link_budget} presents a comparative link budget analysis for LEO satellite communication at L-band frequencies.

\begin{table}[htbp]
\caption{Link Budget Comparison: D2C vs. NTN (LEO, L-band).}
\label{tab:link_budget}
\centering
\begin{tabular}{|l|c|c|c|}
\hline
\textbf{Parameter} & \textbf{Symbol} & \textbf{D2C} & \textbf{NTN} \\
\hline
\makecell[l]{Satellite equivalent \\Isotropically Radiated \\ Power (EIRP) }& $EIRP_{sat}$ & $+52$ dBW & $+45$ dBW \\
\hline
\makecell[l]{Free Space Path Loss \\ (500 km)} & $L_{fs}$ & $-185.5$ dB & $-185.5$ dB \\
\hline
UE Antenna Gain & $G_{rx}$ & $-2$ dBi & $+2$ dBi \\
\hline
Received Power & $P_{rx}$ & $-135.5$ dBW & $-138.5$ dBW \\
\hline
\makecell[l]{Noise Power \\ ($B=200$ kHz)} & $N$ & $-156.3$ dBW & $-156.3$ dBW \\
\hline
Carrier-to-Noise Ratio & $C/N$ & $20.8$ dB & $17.8$ dB \\
\hline
UE Transmit Power & $P_{tx,UE}$ & $+23$ dBm & $+23$ dBm \\
\hline
Uplink $C/N$ & & $11.2$ dB & $14.2$ dB \\
\hline
\textbf{Link Margin (Uplink)} & & $\mathbf{7.2}$ \textbf{dB} & $\mathbf{10.2}$ \textbf{dB} \\
\hline
\end{tabular}
\end{table}

\section{Applications and Use Cases}

\subsection{D2C}

D2C's near-term applications center on emergency SOS, maritime and aviation telemetry, and remote-workforce connectivity. Apple/Globalstar (2022) and T-Mobile/Starlink (2024) demonstrate the emergency use case under the U.S. FCC SCS framework, and 200--300 million D2C-capable handsets may be active by 2027 \cite{computerworld2024apple}. The same legacy-device compatibility also supports low-rate asset tracking and field operations in mining, forestry, and agriculture, although constrained throughput and battery drain remain limiting factors.

\subsection{3GPP NTN}

NTN enables broader use cases spanning global broadband IoT, roaming extension, integrated sensing and communication (ISAC), and high-accuracy localization. It supports container tracking, rail freight, offshore monitoring, and pipeline telemetry with NB-IoT; premium video and alternate backhaul in underserved regions; and Release 19 sensing functions for Earth observation, maritime surveillance, and environmental awareness. Release 18/19 multilateration also improves positioning from roughly 100--200 m to 10--50 m, substantially outperforming D2C localization and extending support to GPS-denied autonomous-vehicle and emergency scenarios \cite{leod2c2025}.

\subsection{Tri-Link}

Within the hybrid model, tri-link architecture primarily targets autonomous driving and resilient corridor deployment. Terrestrial 5G supports real-time perception and cooperative awareness messaging, NTN maintains continuity for map updates and telemetry across rural or maritime gaps, and D2C provides emergency-only fallback for minimum-risk manoeuvres. Release 19 synchronized handover for automotive (SHA) prevents unsafe decisions during transitions, while NTN multi-connectivity can aggregate 50--100 Mbps for HD map refresh and cooperative perception. The same logic extends to trans-oceanic, trans-continental, and polar routes where operator interoperability in Release 20 reduces dependence on a single satellite provider.

Automotive roadmaps already reflect this trend: Tesla pairs D2C SOS with terrestrial connectivity, while Waymo and Mercedes increasingly align with tri-link and NTN-based redundancy concepts for inter-city autonomy and safety-critical V2X.

\section{Security Implications}

\subsection{D2C Vulnerabilities}

D2C deployments exhibit critical security gaps \cite{leod2c2025,d2cdarkside2024}: (1) \textbf{Location Leakage}: Timing advance side-channels leak user location ($\sim 500$–$1000$ m) enabling geolocation tracking of journalists, activists, and high-profile individuals via constellation operator access. (2) \textbf{EMM/IMSI Catcher Attacks}: Satellite-as-rogue-base-station variants exploit Doppler shift masking to bypass IMSI encryption, enabling silent user identification and man-in-the-middle (MITM) attacks; D2C lacks NTN-specific authentication enhancements designed for satellite threat models. (3) \textbf{Constellation Resilience}: Single gateway jamming affects around $30$–$50$ satellites' uplink coverage; beam spoofing induces rogue-satellite handovers; proprietary closed-source implementations complicate vulnerability discovery.

\subsection{NTN Security Advantages vs. D2C Mitigation}

3GPP NTN incorporates security-by-design: NTN-specific authentication, Doppler-robust TMSI allocation, and ISL encryption address satellite threats. Release 19 enforces V2V authentication via time-stamped transponder receipts (preventing replay) and cryptographic signatures; Doppler-robust Global Navigation Satellite Systems (GNSS) augmentation provides spoofing detection for GPS-denied autonomous vehicles \cite{d2cdarkside2024}. D2C hardening measures—timing advance obfuscation (randomized TA reporting), enhanced EMM protection (device-level IMSI encryption), and cryptographic SOS authentication—remain nascent vs. NTN's embedded Release 18 mechanisms. However, D2C enhancements trade-off emergency availability for security.

\subsection{Handover Synchronization in Safety-Critical Moments}

Tri-link transitions must be atomic at the safety-decision level. Release 19 SHA prevents unsafe lane-change decisions during coverage transitions with undefined latency visibility, ensuring atomic state transitions ($< 150$ ms within-tier, $< 500$ ms between-tier) and minimum risk manoeuvre (MRM) pre-computation. D2C explicitly signals higher latency budgets, allowing vehicles to pause overtaking until lower-latency links reestablish.

\section{6G Evolution Pathways}

Release 19 NTN and beyond incorporate technologies that directly address D2C/NTN convergence. AI-driven beam scheduling, predictive handover, and neural-network resource allocation improve discovery latency and spectral efficiency, while new waveforms such as advanced waveforms orthogonal time frequency space (OTFS), affine frequency division multiplexing (AFDM), and orthogonal chirp division multiplexing (OCDM) enhance Doppler resilience in high-mobility LEO channels; fairness and throughput assurance (FTA-NTN) further improves multi-satellite performance by $3$--$5\times$ over single-link operation \cite{3gpp_ts_38_300_2026,6gntnwaveforms2026}.

Release 20 (2028) is expected to formalize the convergence path through seamless subscriber identity module (SIM)-based handover, unified spectrum frameworks, and hybrid terminals that preserve D2C legacy compatibility while enabling NTN performance modes \cite{3gpp_tr_38_895_v20_0_0}. Future satellite platforms may therefore support both transparent payloads for D2C handsets and regenerative gNB functionality for NTN terminals within a common ecosystem.

\section{Conclusion}

D2C and NTN embody distinct engineering trade-offs: D2C prioritizes rapid market entry and legacy device compatibility (forecast: $\sim 100$–$200$ million users by 2027) but faces security vulnerabilities, battery drain, and limited scalability. NTN follows a standards-based evolutionary path (Releases 17–19) delivering superior performance, security-by-design, and 6G-readiness through regenerative payloads and AI optimization (forecast: $500$ million terminals by 2030). Security analysis reveals critical D2C gaps (location leakage, IMSI vulnerabilities) absent in NTN's standardized authentication framework.

The emerging 6G trajectory follows a hybrid convergence roadmap: D2C provides the emergency SOS and resilience layer (leveraging the installed base), while NTN supplies the primary connectivity fabric for broadband, sensing, and critical services. For autonomous driving, a tri-link architecture (terrestrial 5G primary + NTN secondary + D2C tertiary) ensures safety-critical redundancy. Release 20 (2028) formally converges D2C and NTN through seamless handover and unified spectrum frameworks. The future depends not on choosing between them, but on integrating both into a resilient, secure global connectivity ecosystem.

From a deployment perspective, this implies a clear division of roles. Regulators should preserve D2C innovation while enforcing coexistence and baseline security protections, MNOs should position D2C as an emergency and resilience layer rather than a broadband substitute, and automotive OEMs should prioritize latency-aware tri-link designs with authenticated handover and fallback policies. Such a role separation allows D2C to remain an immediate market bridge while NTN matures into the long-term 6G satellite-terrestrial framework.

\section*{Acknowledgment}
The authors acknowledge the financial support by the German Federal Ministry of Research, Technology and Space (BMFTR) within the projects Open6GHub+ {16KIS2406}, X-COM {16KISS007K} and 6G-Coverage {16KIS2425}. However, the authors are solely responsible for the content of the paper. 

\bibliographystyle{IEEEtran}
\bibliography{references}
\end{document}